\documentclass[11pt,a4paper]{article}
\usepackage[hyperref]{acl2020}
\usepackage{times}
\usepackage{latexsym}

\setcitestyle{square}
\usepackage{tablefootnote}
\usepackage{multirow}
\usepackage[symbol]{footmisc}
\usepackage{makecell}

\usepackage[flushleft]{threeparttable}

\setcitestyle{numbers}
\usepackage{microtype}

\aclfinalcopy 

\author{Hyunghoon Cho\footnotemark[1] \\
  \small Broad Institute of MIT and Harvard \\
  \small \texttt{hhcho@broadinstitute.org} \\\And
  Daphne Ippolito\footnotemark[1] \\
  \small University of Pennsylvania \\
  \small \texttt{daphnei@seas.upenn.edu}\\\And
  Yun William Yu\thanks{~~Authors listed alphabetically.} \\
  \small University of Toronto \\
  \small \texttt{ywyu@math.toronto.edu}
  \\}

\date{}

\usepackage{amsmath,amssymb,amsfonts}
\usepackage{hyperref}

\usepackage{algorithmic}
\usepackage{graphicx}
\usepackage{textcomp}
\usepackage{xcolor}
\usepackage{array}
\newcolumntype{L}[1]{>{\raggedright\let\newline\\\arraybackslash\hspace{0pt}}m{#1}}
\usepackage{soul,color}
\soulregister\cite7

\begin{document}

\newcommand{\TODO}[1]{\noindent{\textbf{\textcolor{red}{\{TODO: #1\}}}}}

\title{Contact Tracing Mobile Apps for COVID-19:\\ Privacy Considerations and Related Trade-offs
}



\maketitle

\begin{abstract}
Contact tracing is an essential tool for public health officials and local communities to fight the spread of novel diseases, such as for the COVID-19 pandemic.
The Singaporean government just released a mobile phone app, TraceTogether, that is designed to assist health officials in tracking down exposures after an infected individual is identified.
However, there are important privacy implications of the existence of such tracking apps.
Here, we analyze some of those implications and discuss ways of ameliorating the privacy concerns without decreasing usefulness to public health.
We hope in writing this document to ensure that privacy is a central feature of conversations surrounding mobile contact tracing apps and to encourage community efforts to develop alternative effective solutions with stronger privacy protection for the users.
Importantly, though we discuss potential modifications, this document is not meant as a formal research paper, but instead is a response to some of the privacy characteristics of direct contact tracing apps like TraceTogether and an early-stage Request for Comments to the community.

\textbf{Date written: } 2020-03-24

\textbf{Minor correction: } 2020-03-30
\end{abstract}


\section{Introduction}
The COVID-19 pandemic has spread like wildfire across the globe~\cite{healthmap}.
Very few countries have managed to keep it well-controlled, but one of the key tools that several such countries use is contact tracing~\cite{eames2003contact}.
More specifically, whenever an individual is diagnosed with the coronavirus, every person who had possibly been near that infected individual during the period in which they were contagious is contacted and told to self-quarantine for two weeks~\cite{news_contacttrace}. 
In the early days of the virus, when there were only a few cases, contact tracing could be done manually.
With hundreds to thousands of cases surfacing in some cities, contact tracing has become much more difficult~\cite{news_giveup}.

Countries have been employing a variety of means to enable contact tracing.
In Israel, legislation was passed to allow the government to track the mobile-phone data of people with suspected infection~\cite{bbcnews_israel}.
In South Korea, the government has maintained a public database of known patients, including information about their age, gender, occupation, and travel routes~\cite{washpost_korea}.
In Taiwan, medical institutions were given access to patients’ travel histories~\cite{wang2020response}, and authorities track phone location data for anyone under quarantine~\cite{reuters_taiwan}.
And on March 20, 2020, Singapore released an app that tracks via Bluetooth when two app users have been in close proximity: when a person reports they have been diagnosed with COVID-19, the app allows the Ministry of Health to determine anyone logged to be near them; a human contact tracer can then call those contacts and determine appropriate follow-up actions. 

Solutions that have worked for some countries may not work well in other countries with different societal norms.
We believe that in the United States, in particular, the aforementioned measures are unlikely to be widely adopted.
On the legal side, publicly revealing patients' protected health information (PHI) is a violation of the federal HIPAA Privacy Rule~\cite{hipaa}, and the Fourth Amendment bars the government from requesting phone data without cause~\cite{sc_courts}.
Some of these norms may be suspended during times of crisis---HIPAA has recently been relaxed via enforcement discretion during the crisis to allow for telemedicine~\cite{hipaa_relaxed}, and a public health emergency could well be argued to be a valid cause~\cite{jacobs2011state}.
However, many Americans are wary of sharing location and/or contact data with tech companies or the government, and any privacy concerns could slow adoption of the system~\cite{nytimes_open}.

Singapore's approach of an app, which gives individuals more control over the process, is perhaps the most promising solution for the United States.
However, while Singapore's TraceTogether app protects the privacy of users from each other, it has serious privacy concerns with respect to the government's access to the data.
In this document, we discuss these privacy issues in more detail and introduce approaches for building a contact tracing application with enhanced privacy guarantees, as well as strategies for encouraging rapid and widespread adoption of this system.
We do not make explicit recommendations about how one should build a privacy-preserving contact tracing app, as any design implementation should first be carefully vetted by security, privacy, legal, ethics, and public health experts.
However, we hope to show that there exist options for preserving several different notions of user privacy while still fully serving public health aims through contact tracing apps.

\section{Singapore's TraceTogether App}
On March 20, 2020, the Singaporean Ministry of Health released the TraceTogether app for Android and iOS~\cite{news_tracetogether}.
It operates by exchanging tokens between nearby phones via a Bluetooth connection.
The tokens are also sent to a central server.
These tokens are time-varying random strings, associated with an individual for some amount of time before they are refreshed.
Should an individual be diagnosed with COVID-19, the health officials will ask\footnote{While the health officials ask, it is a crime in Singapore not to assist the Ministry of Health in mapping one's movements, so `ask' is a bit of a misnomer~\cite{tracetogether2020zendesk}.} them to release their data on the app, which includes a list of all the tokens the app has received from nearby phones.
Because the government keeps a database linking tokens to phone numbers and identities, it can  resolve this list of tokens to the users who may have been exposed.

By using time-varying tokens, the app does keep the users private from each other.
A user has no way of knowing who the tokens stored in their app belong to, except by linking them to the time the token was received.
However, the app provides little to no privacy for infected individuals;
after an infected individual is compelled to release their data, the Singaporean government can build a list of all the other people they have been in contact with.
We will formalize these several notions of privacy in Section~\ref{sec:privacy-types}.

\section{Desirable Notions of Privacy}
\label{sec:privacy-types}
Here, we discuss three notions of privacy that are relevant to our analysis of contact-tracing systems: (1) privacy from snoopers, (2) privacy from contacts, and (3) privacy from the authorities.
Note that in this document, we do not rigorously define what it means for information to be private, as this is a topic better left for future works; some popular definitions include information theoretic privacy \cite{shannon1949communication}, k-anonymity \cite{sweeney2002k}, and differential privacy \cite{dwork2006calibrating}.
Furthermore, we discuss only these three notions of privacy to illustrate some of the shortcomings of direct contact-tracing systems. Other recent work has presented a useful taxonomy of the risks and challenges of contact tracing apps \cite{raskar2020apps}.

For any contact tracing app that achieves the aim of telling individuals that they might have been exposed to the virus, there is clearly some amount of information that has to be revealed.
Even if the only information provided is a binary yes/no to exposure, a simple linkage attack \cite{dwork2014algorithmic} can be performed: if the individual was only near to one person in the last two weeks, then there will be an obvious inference about the infection status of that person.
The goal is of course to reduce the amount of information that can be inferred by each of the three parties (snoopers, contacts, and the authorities) while still achieving the public health goal of informing people of potential exposures to help slow the spread of the disease.

Of note, here we use a \textit{semi-honest} model for privacy \cite{goldreich1987solve}, where we do not consider the possibility of malicious actors polluting the database or sending malformed queries, but rather instead just analyze the privacy loss from the information revealed to each party.
A nefarious actor could, for example, falsely claim to be infected to spread panic; this is not a privacy violation, though we do consider this further in the Discussion.
Alternately, when a server exposes a public API, queries can be crafted to reveal more information than intended by the system design, which is indeed a privacy violation.
We leave a more thorough analysis of safeguards for the malicious model to future work.

\subsection{Privacy from Snoopers}
Consider the most na\"{i}ve system for contact tracing, which no reasonable privacy-conscious society would ever use, where the app simply broadcasts the name and phone number of the phone's owner, and nearby phones log this information.
Then, upon diagnosis of COVID-19, the government publishes a public list of those infected, which the app then checks against its list of known recent contacts.
This is clearly problematic as a nefarious passive actor (a `snooper') could track the identities of people walking past them on the street.

A slightly more reasonable system would assign a unique user-ID to each individual, which is instead broadcast out.
This does not have quite as many immediate security implications, though all it would take is a nefarious actor linking each ID to a user before one runs into the same problem, which is known as a `linkage attack.'
Given how easy and common linkage attacks are, this approach also provides insufficient levels of privacy for users~\cite{merener2012theoretical, srivatsa2012deanonymizing}.

The Singaporean app TraceTogether does better, in that it instead broadcasts random time-varying tokens as temporary IDs.
Because these tokens are random and change over time, someone scanning the tokens while walking down the street will not be able to track specific users across different time points, as their tokens are constantly refreshed.
Note that the length of time before refreshing a token is an important parameter of the system (too infrequent and users can still be tracked, too frequent and the amount of tokens that need to be stored by the server could be huge), but with a reasonable refresh rate, the users are largely protected against attacks by snoopers in public spaces.

\subsection{Privacy from Contacts}
Here, the term \emph{contact} is defined as any individual with whom a user has exchanged tokens in the contact tracing app based on some notion of physical proximity.
Privacy from contacts is harder to achieve, because the information that needs to be passed along is whether one of the individual's contacts has been diagnosed with COVID-19, so some information has to be revealed.

The TraceTogether app gives privacy from contacts by instead putting trust in government authorities.
When TraceTogether alerts a contact that they have been exposed to COVID-19, the information comes directly from the Singaporean Ministry of Health, and no additional information is shared (to our knowledge) that could identify the individual that was diagnosed.
Thus, TraceTogether does protect users' privacy from each other, except for what can be inferred based on the user's full list of contacts, as the only information that is revealed to the user is a binary exposure indicator, which is arguably the minimum possible information release for the system to be useful.

\subsection{Privacy from the Authorities}
Protecting the privacy of the users from the authorities, i.e.~whoever is administering the app, whether that is a government agency or a large tech company, is also a challenging task.
Clearly, in the absence of a fully decentralized peer-to-peer system, any information sharing among phones with the app installed will have to be mediated by some coordinating servers.
Without any protective measures (e.g.~based on cryptography), the coordinating servers are given an inordinate amount of knowledge.

TraceTogether does not privilege this type of privacy, instead making use of relatively high trust in the government in its design.
While it does not deliberately gather more information than necessary to build a contact map---for example, it does not use GPS location information, as Bluetooth is sufficient for finding contacts---it also does not try to hide anything from the Singaporean government.
When a user is diagnosed with COVID-19 and gives their list of tokens to the Ministry of Health, the government can retrieve the mobile numbers of all individuals that user has been in contact with.
Thus, neither the diagnosed user, nor the exposed contacts, have any privacy from the government.

Furthermore, because the government maintains a database linking together time-varying tokens with mobile numbers, they can also, in theory, track people's activities without GPS simply by placing Bluetooth receivers in public places.
There is no reason to disbelieve the TraceTogether team when they state that they do not attempt to track people's movements directly; however, the data they have could be employed to do so.
Citizens of countries such as the U.S. trust authorities much less than Singaporeans~\cite{barometer2019january}, so the privacy trade-offs that Singaporeans are willing to make may not be the same ones that Americans will accept.

\newcolumntype{?}{!{\vrule width 1pt}}
\newcolumntype{N}{>{\centering\arraybackslash}m{.5in}}

\begin{table*}[tbp]
\small
\begin{threeparttable}
    \caption{\textbf{Comparison of contact tracing systems discussed in this document with respect to privacy of the users in the semi-honest model and required computational infrastructure.}
    }

\begin{tabular}{L{1.4cm}?L{1.1cm}|L{1.1cm}|L{2.0cm}|L{2.5cm}|L{2.5cm}|L{1.6cm}}
\Xhline{2\arrayrulewidth}
 & \multirow{2}{1.1cm}{Privacy from snoopers} & \multicolumn{2}{p{3.1cm}|}{\centering Privacy from contacts}     & \multicolumn{2}{p{5cm}|}{\centering Privacy from authorities}   & \multirow{2}{1.6cm}{Infrastructure requirements}     \\ 
\cline{3-6}
&      & Exposed user & Diagnosed user   & Exposed user    & Diagnosed user    &       \\
\Xhline{2\arrayrulewidth}
Trace Together~\cite{news_tracetogether}            & Yes   & Yes    & Yes     & No. Exposure status and all tokens revealed.      & No. Infection status, all tokens, and all contact tokens revealed.                      & Minimal                          \\ \hline
Polling-based\footnotemark[1] (\S\ref{subsec:pollingonly})            & Yes                                    & Yes    & Yes\footnotemark[2]  & Partial. Susceptible to linkage attacks.         & Partial. Susceptible to linkage attacks.                                                & Low. Single server.     \\ \hline
Polling-based with mixing (\S\ref{section:pollingmixing}) & Yes                                    & Yes      & Yes\footnotemark[2]  & Almost private. Protects against linkage attacks by mixing tokens from different users. & Almost private. Protects against linkage attacks by mixing tokens from different users. & Medium. Multiple servers for mixing.      \\ \hline
Public database (\S\ref{subsec:public})           & Yes                                    & Yes                & Partial. Info leaked at time of token exchange.                                                                                       & Yes                                                                                     & Partial. Susceptible to linkage attacks.                                                & Communication cost to phones is high. \\ \hline
Private messaging system (\S\ref{sec:pms})  & Yes                                    & Yes          & Partial. Info leaked at time of token exchange. \footnotemark[3]                  & Yes      & Yes              & High. Multiple servers performing crypto.
\\
\Xhline{2\arrayrulewidth}
\end{tabular}
  \begin{tablenotes}
      \small
      \item \footnotemark[1] Augmenting with random tokens does not improve privacy.
      \item \footnotemark[2] However, if contacts are malicious, and they send malformed queries (e.g. a query that includes only a single token), the diagnosed individual only has the same privacy level as in the public database solution.
      Namely, there's only partial privacy because information is leaked through knowing the time of token exchange.
      \item \footnotemark[3] This information leakage might be fixable using data aggregation based on multi-key homomorphic encryption, but we do not do so here.
    \end{tablenotes}
    \label{tab:my_label}
\end{threeparttable}
\end{table*}

\section{Privacy-Enhancing Augmentations to the TraceTogether System}
\label{section:augmentations}
Here, we discuss potential approaches to build upon the TraceTogether model to obtain a contact tracing system with differing privacy characteristics for the users.
Though important and highly nontrivial, various technical and engineering challenges behind the exchange of Bluetooth tokens~\cite{tracetogether2020zendesk2} are outside the scope of this document.
Our abstraction is that there exists some mechanism for nearby phones to exchange short tokens if the devices come within 6 feet of each other---the estimated radius within which viral transmission is a considerable risk~\cite{cdc_spread}.
We are primarily concerned with the construction of those tokens, and how those tokens can be used to perform contact tracing in a privacy-preserving manner.

First, we formally describe the TraceTogether system.
Let Alice and Bob be users of the app, and let Grace be the government server (or other central authority).
Alice generates a series of random tokens $A = \{a_0, a_1, \ldots \}$, one for each time interval, and Bob generates a similar series of tokens $B = \{b_0, b_1, \ldots \}$, all drawn randomly from some space $\{0, 1\}^N$.
They also both report their list of tokens $A$ and $B$, as well as their phone numbers to Grace.
At a time $t$, Alice and Bob encounter each other, exchanging $a_t$ and $b_t$.
Alice and Bob keep lists of contact tokens $\hat{A} = \{ \hat{a}_0, \hat{a}_1, \ldots \}$ and $\hat{B} = \{ \hat{b}_0, \hat{b}_1, \ldots \}$ respectively. These consist of tokens from every person they were exposed to; i.e.~$b_t \in \hat{A}$ and $a_t \in \hat{B}$ because Alice and Bob exchanged tokens at time $t$.
Five days later, Bob is diagnosed with COVID-19, and sends his list of contact tokens $\hat{B}$, which includes $a_t$, to Grace.
Grace then matches each $\hat{b}_i$ to a phone number, reaches out to those individuals, including Alice, and advises them to quarantine themselves because they may have been exposed to the virus.

\subsection{Partially Anonymizing via Polling}
\label{subsec:pollingonly}
Instead of having Grace reach out to Alice when Bob reports that he has been diagnosed, a more privacy-conscious alternative is for Alice to ``poll'' Grace on a regular basis.
In this setting, Grace maintains the full database, and Alice asks Grace if she has been exposed.
This alternative does not require Alice and Bob to send their phone numbers to Grace.
In this setting, there are two reporting choices for when Bob wishes to declare his diagnosis of COVID-19.
Bob can send his own tokens $B$ to Grace, or he can send the contact tokens $\hat{B}$ to Grace.
In the former case, Alice needs to send Grace her contact tokens $\hat{A}$ to see if any have been diagnosed with COVID-19.
In the latter case, Alice needs to send Grace her own tokens $A$ to ask if any of them have been published.
Either way, Grace is able to inform Alice that she has been exposed, without revealing Bob's identity.
This presupposes that Alice is Honest but Curious (semi-honest); if Alice is malicious and crafts a malformed query containing only the token she exchanged with Bob, she may be able to reveal Bob's identity.

Note that in either version of this system, individuals still have privacy from snoopers and from contacts.
However, they additionally gain some amount of privacy from authority, as Grace does not have their mobile numbers.
Of course, Grace does have some ability to perform linkage attacks.
If Bob publishes to Grace his own tokens $B$ upon being diagnosed, and Alice queries Grace with all her contact tokens $\hat{A}$, then Grace can attempt to link those sets of tokens to individuals or geographic areas; further, Grace can also monitor the source of Alice and Bob's queries (i.e.~IP addresses of phones).
For example, if Grace has Bluetooth sensors set up in public places, she can then trace Alice and Bob's geographic movements.
That kind of location trace is often sufficient to deanonymize personal identities~\cite{srivatsa2012deanonymizing}.
Alternatively, the same is true if Bob publishes his contact tokens to Grace and Alice queries Grace with her own tokens.
Thus, there is not perfect privacy from the authorities, but still better than in the original TraceTogether system, at the cost of potentially lower privacy for Bob in the malicious model.

\subsection{Ineffectiveness of Adding Spurious Tokens for Further Anonymization}
To further anonymize the polling-based system to increase privacy from authorities, there are a number of techniques that can be used to hide Alice and Bob's identities.
Let's begin with a simple approach---that doesn't actually work---to give some intuition before moving on to more effective approaches.
Consider injecting random noise by augmenting the data with artificial tokens.
Whenever Alice and Bob send information to Grace (either in the form of a diagnosis report or a query), they can augment their tokens with random ones.
Note that some care has to be taken in deciding which distribution to draw the random tokens from.
Not only should the system keep the probability of spurious matches low, but the distributions should also be designed to make inferences by Grace difficult.

For example, assume that Alice and Bob sample their tokens uniformly at random from $\{0, 1\}^{N}$, where
$N$ is chosen to be sufficiently large that accidental collisions between individuals' tokens are unlikely.
Suppose Bob sends to Grace his own tokens $B$ upon being diagnosed, and Alice queries Grace with all her contact tokens $\hat{A}$.
In theory, Bob could augment his own tokens with a set of $n$ random tokens $\{r_i\}_{i=1}^n$ drawn uniformly from $\{0, 1\}^{N}$, and send those to Grace as well.
Unfortunately, $N$ was chosen to prevent accidental collisions; this means that the probability that the additional random tokens correspond to the tokens broadcast by any individual is vanishing small.
But then, there is actually little to no privacy gained.
Grace can just assume that the augmented set of tokens correspond to Bob, and perform the same linkage analysis that she would with only the correct set of tokens.
This does nothing but pollute Grace's database with extra data, without affording any real privacy gains for Bob.
Similarly, Alice also cannot obfuscate her exposure through Bob from Grace, because any extra tokens she sends to Grace will not change the fact that she has Bob's token as one of her contacts.

The root of the problem is that Grace has access to the universe of all tokens through user queries, and so can simply filter out all of the random tokens generated.
Thus, random noise is ineffective for hiding information from Grace.

\subsection{Enhancing Anonymity by Mixing Different Users' Tokens}
\label{section:pollingmixing}
Although introducing spurious random tokens into the system achieves little in terms of privacy, as discussed in the previous subsection, a slight modification of this idea leads to meaningful privacy guarantees.
The issue is that Grace has access to the entire universe of tokens, as well as both of the sets of tokens corresponding to Alice and Bob, possibly augmented with random noise.
Instead of hiding true tokens with random noise, suppose the system includes a set of $M$ honest-but-curious non-colluding ``mixing'' servers not controlled by Grace that aggregate data before forwarding it on to Grace.

When Bob is diagnosed with COVID-19, he partitions the tokens he wishes to send (depending on the setup of the system, either his own tokens, or those of his contacts) into $M$ groups, and sends each group to one of the mixing servers.
The mixing servers then combine Bob's data with that of other users diagnosed with COVID-19 before forwarding it onto Grace.
Similarly, Alice does the same thing for querying, except she also needs to wait on a response from the mixing server for each of the tokens she sends.
The linkage problem then becomes much more difficult for Grace, because the valid tokens for individuals have been split up.
Similarly, each mixing server only has access to a subset of the tokens corresponding to each individual, making the linkage analysis more difficult for them.
Of course, if the mixing servers collude, then the privacy reduces to that of the standard polling-based approach.

Note that this approach can also be simulated without the mixing servers by either Alice or Bob if they have access to a large number of distinct IP addresses.
They can simply send their queries and tokens with some time delay from the different IP addresses, preventing Grace from linking all of them together.
However, this approach may not be feasible for most users.

\subsection{Public Database of Infected Users' Tokens is Efficient but Less Private}
\label{subsec:public}
Alternatively, Grace can simply publish the entire database of tokens she receives from infected individuals, including the ones from Bob.
If Alice simply downloads the entire database, and locally queries against it, then no information about Alice's identity is leaked to Grace.

This approach may seem less computationally feasible, especially on mobile devices.
In circumstances where the total number of people infected is not very high, this approach works, as evidenced by the South Korean model~\cite{washpost_korea}, though the approach may fail as the epidemic reaches a peak.
However, the computational and transmission cost can be partially ameliorated by batching together Grace's database, so that Alice is not downloading the entire thing.
For example, in the version where Bob sends his own tokens $B$ to Grace, Alice can download batches corresponding to her contact tokens $\hat{A}$.
If each batch has e.g.~50 tokens, then Grace does not know which of those 50 tokens Alice came into contact with.

Unfortunately, it is worth noting that this approach decreases Bob's privacy from Alice, because Alice knows when she encountered the token Bob sent; she can then limit the number of possible individuals who could have sent the token based on who she was in contact with during the time she encountered Bob's token.
If the token she exchanged with Bob is present in the database, she gets a hint as to the disease status of one of the individuals she was in contact with during the token exchange.

\begin{figure*}
    \centering
    \includegraphics[width=\textwidth]{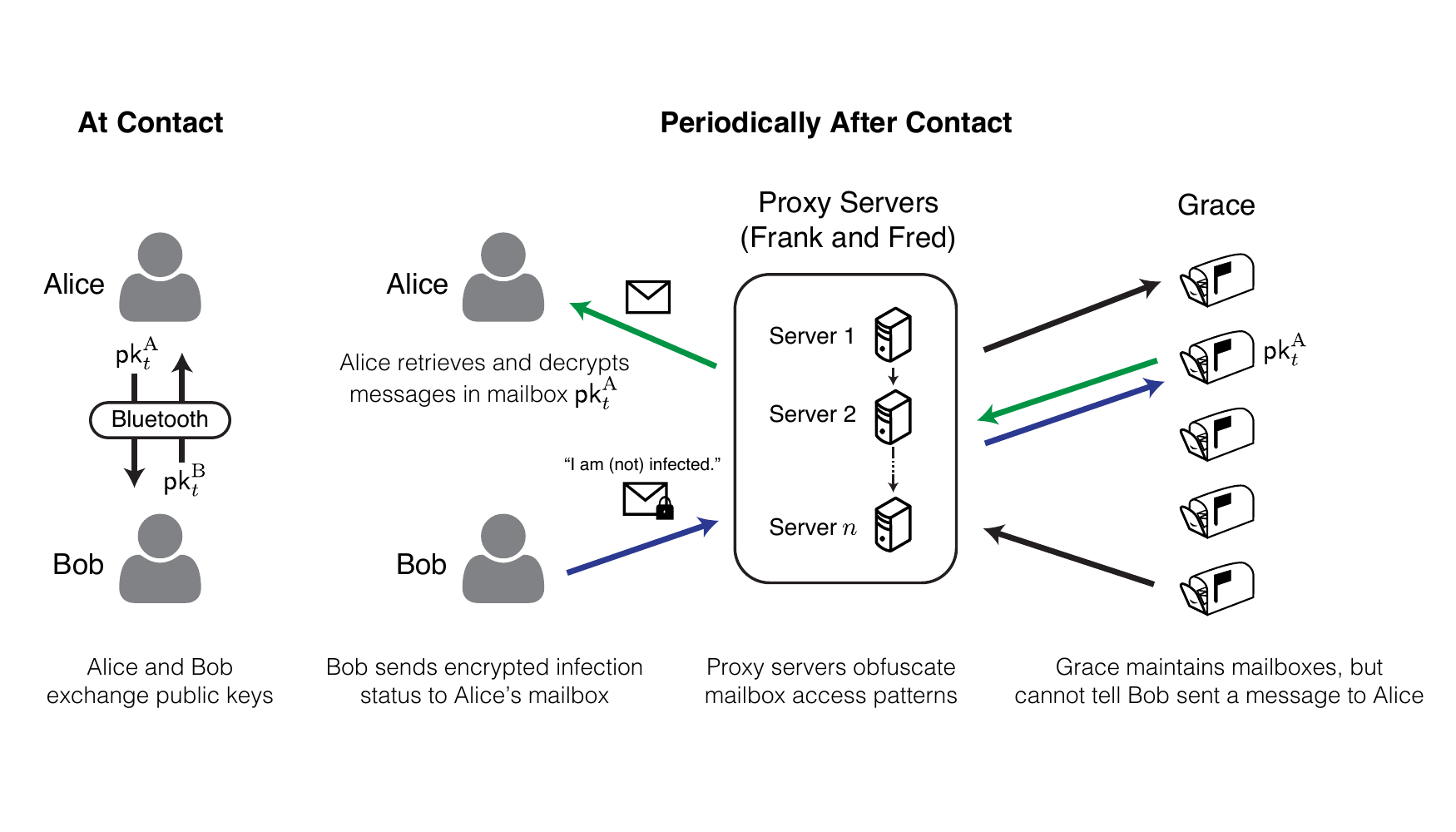}
    \vspace{-4em}
    \caption{\textbf{Overview of contact tracing based on private messaging systems.} When Alice and Bob are near each other they exchange public keys as tokens. They then periodically encrypt (using each other's public key, followed by the public keys of the proxy servers) a message indicating their infection status, and send it to the proxy server. They also periodically query the proxy server for messages posted to the mailboxes corresponding to their public keys to find out whether they have been exposed to the virus.}
    \label{fig:diagram}
\end{figure*}

\section{Privacy from Authorities based on Private Messaging Systems}
\label{sec:pms}
None of the easy-to-implement augmentation ideas given in Section~\ref{section:augmentations} guarantee full privacy from the authorities.
At a cost of more computation, however, we believe that a solution for secure contact tracing can be built using modern cryptographic protocols.
In particular, private messaging systems \cite{van15, tyagi2017stadium, corrigan15} and private set intersection (cardinality) \cite{freedman2004efficient, kissner2005privacy, de2010practical, de2012fast} protocols seem especially relevant.
The sketch we provide below is based on private messaging systems, though we do not claim this to be an optimal implementation.

We will give the intuition here before going into technical details necessary for an effective implementation.
First, we replace the random tokens $(a_t,b_t)$ exchanged by Alice and Bob with random public keys $(\text{pk}^A_t,\text{pk}^B_t)$ from asymmetric encryption schemes \cite{simmons1979symmetric}.
The matching secret keys are stored locally on each of Alice's and Bob's phones.
Then, imagine that Grace has established a collection of mailboxes, one for each public key that Alice and Bob exchange.
Additionally, we introduce Frank and Fred. Frank \textit{forwards} messages to/from Fred. Fred forwards messages to/from Grace. They do not tell each other the source of the messages.
At fixed time points after Bob's contact with Alice (up to some number of days), Bob addresses a message to Alice encrypted using the public key Alice gave Bob.
Bob gives the message to Frank, who then forwards it on to Grace (through Fred), who puts it in Alice's mailbox.
The content of the message is Bob's current infection status, and the reason he sends messages at fixed time points is to prevent Frank from figuring out Bob's infection status from the fact that he is sending messages.
Alice checks all of the mailboxes corresponding to her last several days worth of broadcasted public keys.
In one of the mailboxes, she then receives and decrypts Bob's message, and learns whether she has been exposed to the virus.
Grace cannot decrypt the message Bob sends to Alice because it is protected by asymmetric encryption.
Furthermore, to protect Alice's privacy, she can also access her mailboxes through Frank and Fred, who deliver the messages in Alice's mailboxes to her without revealing which mailboxes she owns.

Contact tracing can be viewed as a problem of secure communication between pairs of users who came into contact in the physical world.
The communication patterns of who is sending messages to whom can reveal each individual’s contact history to the service provider (Grace).
This notion is known as metadata privacy leakage in computer security~\cite{greschbach12}, where the metadata associated with a message (e.g. sender/recipient and time) is considered sensitive, in addition to the actual message contents.
In the contact tracing case, such metadata could reveal who has been in contact with whom, potentially revealing the users' sensitive activities.
We believe that recent technical advances~\cite{kwon20,van15,corrigan15} for designing scalable private messaging systems with metadata privacy present a promising path for developing a similar platform for secure contact tracing.

Following recent works, our idea is to leverage a `mix network’~\cite{chaum81}, which is a routing protocol that uses a chain of proxy servers (Frank/Fred) that individually shuffle the incoming messages before passing them onto the next server, thereby decoupling the sender of each message from its destination---these types of mix networks are perhaps most well-known for being the basis of the Onion Router/Tor anonymity network \cite{reed1998anonymous}.
This is a more sophisticated use of mixing servers than described in Section~\ref{section:pollingmixing} for the polling based solution.
When Bob wishes to send his encrypted message to Alice, he first encrypts it multiple times with public keys corresponding to each of the servers in the mix network.
Because the messages are encrypted in multiple layers, and each server peels only the outermost layer, the final destination (Alice's mailbox) is revealed only to the last server, and only Alice can read the content of the message (i.e.~infection status).
To prevent Grace from learning the identity associated with each mailbox, Alice can also access her mailboxes through the mix network, which shuffles the traffic to decouple the mailboxes from their owners.
As long as one of the servers is neither breached nor controlled by the adversary, the final message cannot be linked to a specific sender even if the adversary has full control of the rest of the network.
Such a system for private communication could allow the users (Bob) to share their infection status with their recent contacts (Alice) while hiding the metadata of their contact patterns from the service providers.
The involvement of non-government entities, such as an academic institution or a hospital, in the mix network may help increase users’ trust in the system and lower the bar for adoption.

There are several remaining issues that will need to be addressed for this system to be widely adopted.
First, if time-varying IDs are used, then the user receiving a token from a nearby person could infer the identity of the sender based on their travel history; i.e.~Alice might be able to infer who Bob is based on the time they exchanged the tokens, as described in Section~\ref{subsec:public} in the case where the database is made public.
This loss of privacy from contacts can be partially alleviated by choosing a less frequent token refresh, so that with high likelihood, Alice cannot completely identify Bob by the time interval. 
Actual implementations much decide on the right tradeoffs between Alice and Bob's privacy from eachother and authorities, as well as contact tracing effectiveness.
Another possible way to mitigate this problem would be to aggregate the messages for Alice on the server before making the results available to her.
The messages are encrypted under different public keys, but it may be possible to use multi-key homomorphic encryption schemes~\cite{lopez2012fly,chen2019efficient} which allow computation over ciphertexts encrypted with different public keys to sum up the count of `infected' messages.
We defer the details of approach to future work.

One other issue is that the volume of messages delivered to each user may reveal how socially active each user has been, which could be considered sensitive by some users.
Approaches to flatten the distribution with dummy messages could alleviate this concern.
Flattening the distribution with dummy messages may however lead to scalability challenges for existing private messaging systems.
Though many techniques~\cite{kwon20,van15,corrigan15} have been proposed to address this challenge, further discussion among the stakeholders is needed to determine the suitable trade-off between the level of latency that can be tolerated and the level of privacy guarantees desired by the users.
Ultimately, though, private messaging systems enable provable privacy from the authorities while still maintaining the usefulness of contact tracing.

\section{Strategies for Encouraging Widespread Adoption}
Contact tracing apps depend on the network effect and critical mass to work.
Having the app go `viral' requires that people trust the app enough to install it and are enthusiastic enough to convince their friends to do the same.
After all, app adoption must have a higher `transmission rate' than the virus itself in order for it to be effective. 
Providing strong privacy guarantees would likely encourage voluntary adoption.
Any app needs to clearly explain privacy guarantees in ways understandable by the average user, which was our motivation in describing here the different types of privacy (from snoopers, contacts, and the authorities) that the app should be able to provide to users in order to earn their trust.

On that note, we believe it is imperative for any app to be open source and audited by both security professionals and privacy advocates.
This is not yet true for TraceTogether, but the app's creators do claim that they will release the source code soon~\cite{news_opensource}.
Furthermore, open sourcing allows different countries to customize such apps for their particular use cases and cultural preferences.

Also, while in some countries it may be difficult to enforce a government mandate that all residents install an app, it is possible to have this as a requirement for entering certain public places.
Such a practice has precedence in so-called implied consent laws, such as agreeing to field sobriety tests when getting a driver's license~\cite{wagenaar1995methods}.
One could imagine grocery stores, schools, and universities requiring installing a contact tracing app as a precondition for entrance.
This does not stop users from uninstalling or turning off the app off-premises, but it would at least be useful in getting people over the initial activation barrier of installation.

Finally, some amount of social pressure may also assist in reaching widespread adoption.
Contact tracing apps, by design, know how many other people close by have the app installed.
An app could display that number.
Given this knowledge, a user may be incentivized to attempt to persuade others nearby to install the app, in the interest of public health.

\section{Discussion}
In this document, we discuss ways to build an app for contact tracing, based upon the premise that phones can broadcast tokens to all nearby phones.
Notably, we do not address the engineering behind applying Bluetooth to enable such a feature.
Nor do we address the possibility of location data collection for assisting epidemiologists in forecasting disease spread \cite{pei2018forecasting}.
We also do not discuss appropriate selection of token refresh interval and frequency at which phones should poll for nearby ones, which are important factors for balancing privacy and efficiency---stale IDs have been seen to permit linkage attacks in other similar contexts \cite{sarma2002rfid}.
Lastly, we also do not build a full model for privacy of contact tracing, which is a delicate and easy-to-get-wrong task that requires much more careful research.
Instead, we focus only on the privacy implications of a dedicated contact tracing app, in the hopes that providing sufficiently strong privacy guarantees would assist an app in gaining the critical mass needed to be effective.

Note that here we only discuss direct contact tracing using Bluetooth proximity networks, without using any location data.
Some indirect proposals for contact tracing instead simply securely log the user's location history, which is then given to the authorities if a user is diagnosed with COVID-19~\cite{safekit}.
This approach has the benefit of not requiring network effects, because single individuals can track their locations without needing their contacts to have the app.
The approach of logging location history is inherently less private than direct contact tracing, but that may possibly be resolved with appropriate safeguards and redactions~\cite{safekit}.
Furthermore, hybrid approaches involving both GPS data and Bluetooth proximity networks may prove to be useful to public health officials in modelling disease spread beyond just contact tracing \cite{covidwatch}.

We first discussed how, with just minor modifications, a polling-based direct contact tracing solution allows for some anonymity from authorities, which is lacking in the Singaporean Ministry of Health app TraceTogether.
We believe that this may help an app succeed in countries such as the U.S., where many citizens are loath to give too much data to the government.

Even the polling-based solution still reveals quite a bit of information to the authorities, who could make use of linkage analysis to track individual users.
However, utilizing additional mixing servers is relatively practical and does provide additional protection.
Alternately, a system can follow the South Korean model of openly publishing data about patients diagnosed with COVID-19, trading off some of their privacy to enhance the privacy of individuals who are trying to determine if they have been exposed.

However, if we are willing to invest in additional computational resources, it is possible to achieve increased privacy from snoopers, contacts, and the authorities, and we propose the beginnings of one approach using private messaging systems, which we hope will be further expanded upon in future works.
This is more computationally expensive, but would assure users that they do not have to give up their privacy in order to take part in public contact tracing efforts.
Indeed, the chief selling point would be that they would get additional information on their exposure without needing to trust any individual third party with their private location or medical information.
We believe that such a guarantee would go a long way towards mass adoption of a contact tracing app in the United States.

Future work remains to actually build such an app, of course, and additional engineering, security, and policy considerations are sure to arise.
For example, scalability of the data structures used in the servers may become a major issue when the number of infected individuals rises.
One additional concern which we have not addressed is that of nefarious actors seeking to spread panic by falsely claiming to be infected.
This could be prevented by allowing only hospital workers to trigger the broadcast of infection status, as in Singapore's system, where the Ministry of Health directly contacts those exposed, though that of course trades away some of the privacy of diagnosed patients.
Alternately, others have proposed cryptographic verification of contact events, which could perhaps be extended to infection event broadcast without giving direct access of tokens to the authorities \cite{csct}.
However, given that some cities are already rationing testing kits and doctors' visits to only the most serious cases~\cite{latimes_limitedtesting, philinq_limitedtesting}, restricting self-reporting might result in many instances of virus spread to be missed.
Alternately, the system can also be designed to separate self-reports from confirmed reports by simply keeping two databases.

Our goal in writing this document is to start a conversation on (1) what kinds of privacy trade-offs people are willing to endure for the sake of public health, and (2) the fact that with sufficient computational resources and use of cryptographic protocols, app-based contact tracing can be accomplished without completely sacrificing privacy.
Because bad early design choices can persist long after roll-out, we hope that developers and policy-makers will give privacy considerations careful thought when designing new contact tracing apps.

\section*{Acknowledgment}
\addcontentsline{toc}{section}{Acknowledgment}
We would like to thank David Rolnick, Adam Sealfon, Noah Daniels, and Michael Wirth for helpful comments.

\bibliographystyle{./bibliography/IEEEtran}
\bibliography{./bibliography/IEEEexample}

\begin{thebibliography}{10}
\providecommand{\url}[1]{#1}
\csname url@samestyle\endcsname
\providecommand{\newblock}{\relax}
\providecommand{\bibinfo}[2]{#2}
\providecommand{\BIBentrySTDinterwordspacing}{\spaceskip=0pt\relax}
\providecommand{\BIBentryALTinterwordstretchfactor}{4}
\providecommand{\BIBentryALTinterwordspacing}{\spaceskip=\fontdimen2\font plus
\BIBentryALTinterwordstretchfactor\fontdimen3\font minus
  \fontdimen4\font\relax}
\providecommand{\BIBforeignlanguage}[2]{{%
\expandafter\ifx\csname l@#1\endcsname\relax
\typeout{** WARNING: IEEEtran.bst: No hyphenation pattern has been}%
\typeout{** loaded for the language `#1'. Using the pattern for}%
\typeout{** the default language instead.}%
\else
\language=\csname l@#1\endcsname
\fi
#2}}
\providecommand{\BIBdecl}{\relax}
\BIBdecl

\bibitem{healthmap}
\BIBentryALTinterwordspacing
``{Novel Coronavirus Map from HealthMap},'' March 2020. [Online]. Available:
  \url{https://www.healthmap.org/covid-19/}
\BIBentrySTDinterwordspacing

\bibitem{eames2003contact}
K.~T. Eames and M.~J. Keeling, ``{Contact tracing and disease control},''
  \emph{Proceedings of the Royal Society of London. Series B: Biological
  Sciences}, vol. 270, no. 1533, pp. 2565--2571, 2003.

\bibitem{news_contacttrace}
D.~Normile, ``{Coronavirus cases have dropped sharply in South Korea. What’s
  the secret to its success?}''
  \url{https://www.sciencemag.org/news/2020/03/coronavirus-cases-have-dropped-sharply-south-korea-whats-secret-its-success},
  2020, accessed: 2020-03-23.

\bibitem{news_giveup}
B.~Chappell, ``{Coronavirus: Sacramento County Gives Up On Automatic 14-Day
  Quarantines},''
  \url{https://www.npr.org/sections/health-shots/2020/03/10/813990993/coronavirus-sacramento-county-gives-up-on-automatic-14-day-quarantines},
  2020, accessed: 2020-03-23.

\bibitem{bbcnews_israel}
\BIBentryALTinterwordspacing
J.~Tidy, ``{Coronavirus: Israel enables emergency spy powers},'' \emph{BBC
  News}, March 2020. [Online]. Available:
  \url{https://www.bbc.com/news/technology-51930681}
\BIBentrySTDinterwordspacing

\bibitem{washpost_korea}
\BIBentryALTinterwordspacing
M.~J. Kim and S.~Denyer, ``{A ‘travel log’ of the times in South Korea:
  Mapping the movements of coronavirus carriers },'' \emph{The Washington
  Post}, March 2020. [Online]. Available:
  \url{https://www.washingtonpost.com/world/asia\_pacific/coronavirus-south-korea-tracking-apps/2020/03/13/2bed568e-5fac-11ea-ac50-18701e14e06d\_story.html}
\BIBentrySTDinterwordspacing

\bibitem{wang2020response}
C.~J. Wang, C.~Y. Ng, and R.~H. Brook, ``{Response to COVID-19 in Taiwan: Big
  Data Analytics, New Technology, and Proactive Testing},'' \emph{JAMA}, 2020.

\bibitem{reuters_taiwan}
\BIBentryALTinterwordspacing
Y.~Lee, ``{Taiwan's new 'electronic fence' for quarantines leads wave of virus
  monitoring},'' March 2020. [Online]. Available:
  \url{https://www.reuters.com/article/us-health-coronavirus-taiwan-surveillanc-idUSKBN2170SK}
\BIBentrySTDinterwordspacing

\bibitem{hipaa}
\BIBentryALTinterwordspacing
``{HIPAA Privacy Rule},'' December 2000. [Online]. Available:
  \url{https://www.hhs.gov/hipaa/for-professionals/privacy/index.html}
\BIBentrySTDinterwordspacing

\bibitem{sc_courts}
\BIBentryALTinterwordspacing
C.~J. Roberts, ``{Carpenter v. United States},'' \emph{Supreme Court of the
  United States}, no. 16-402, 2018. [Online]. Available:
  \url{https://www.supremecourt.gov/opinions/17pdf/16-402\_h315.pdf}
\BIBentrySTDinterwordspacing

\bibitem{hipaa_relaxed}
\BIBentryALTinterwordspacing
``{Notification of Enforcement Discretion for telehealth remote communications
  during the COVID-19 nationwide public health emergency},'' March 2020.
  [Online]. Available:
  \url{https://www.hhs.gov/hipaa/for-professionals/special-topics/emergency-preparedness/notification-enforcement-discretion-telehealth/index.html}
\BIBentrySTDinterwordspacing

\bibitem{jacobs2011state}
A.~J. Jacobs, ``{Is state power to protect health compatible with substantive
  due process rights},'' \emph{Annals Health L.}, vol.~20, p. 113, 2011.

\bibitem{nytimes_open}
\BIBentryALTinterwordspacing
R.~Pérez-Peña, ``{Virus Hits Europe Harder Than China. Is That the Price of
  an Open Society? },'' \emph{New York Times}, March 2020. [Online]. Available:
  \url{https://www.nytimes.com/2020/03/19/world/europe/europe-china-coronavirus.html}
\BIBentrySTDinterwordspacing

\bibitem{news_tracetogether}
\BIBentryALTinterwordspacing
``{Help speed up contact tracing with TraceTogether},'' \emph{Singapore
  Government Blog}, March 2020. [Online]. Available:
  \url{https://www.gov.sg/article/help-speed-up-contact-tracing-with-tracetogether}
\BIBentrySTDinterwordspacing

\bibitem{tracetogether2020zendesk}
T.~TraceTogether, ``{Can I say no to uploading my TraceTogether data when
  contacted by the Ministry of Health?}''
  \url{https://tracetogether.zendesk.com/hc/en-sg/articles/360044860414-Can-I-say-no-to-uploading-my-TraceTogether-data-when-contacted-by-the-Ministry-of-Health-},
  2020, accessed: 2020-03-23.

\bibitem{shannon1949communication}
C.~E. Shannon, ``Communication theory of secrecy systems,'' \emph{Bell system
  technical journal}, vol.~28, no.~4, pp. 656--715, 1949.

\bibitem{sweeney2002k}
L.~Sweeney, ``k-anonymity: A model for protecting privacy,''
  \emph{International Journal of Uncertainty, Fuzziness and Knowledge-Based
  Systems}, vol.~10, no.~05, pp. 557--570, 2002.

\bibitem{dwork2006calibrating}
C.~Dwork, F.~McSherry, K.~Nissim, and A.~Smith, ``Calibrating noise to
  sensitivity in private data analysis,'' in \emph{Theory of cryptography
  conference}.\hskip 1em plus 0.5em minus 0.4em\relax Springer, 2006, pp.
  265--284.

\bibitem{raskar2020apps}
R.~Raskar, I.~Schunemann, R.~Barbar, K.~Vilcans, J.~Gray, P.~Vepakomma,
  S.~Kapa, A.~Nuzzo, R.~Gupta, A.~Berke \emph{et~al.}, ``Apps gone rogue:
  Maintaining personal privacy in an epidemic,'' \emph{arXiv preprint
  arXiv:2003.08567}, 2020.

\bibitem{dwork2014algorithmic}
C.~Dwork, A.~Roth \emph{et~al.}, ``The algorithmic foundations of differential
  privacy,'' \emph{Foundations and Trends{\textregistered} in Theoretical
  Computer Science}, vol.~9, no. 3--4, pp. 211--407, 2014.

\bibitem{goldreich1987solve}
O.~Goldreich, S.~Micali, and A.~Wigderson, ``How to solve any protocol
  problem,'' in \emph{Proc. of STOC}, 1987.

\bibitem{merener2012theoretical}
M.~M. Merener, ``{Theoretical results on de-anonymization via linkage
  attacks},'' \emph{Transactions on Data Privacy}, vol.~5, no.~2, pp. 377--402,
  2012.

\bibitem{srivatsa2012deanonymizing}
M.~Srivatsa and M.~Hicks, ``{Deanonymizing mobility traces: Using social
  network as a side-channel},'' in \emph{Proceedings of the 2012 ACM conference
  on Computer and communications security}, 2012, pp. 628--637.

\bibitem{barometer2019january}
\BIBentryALTinterwordspacing
E.~T. Barometer, ``January 20, 2019,'' 2019. [Online]. Available:
  \url{https://www.edelman.com/sites/g/files/aatuss191/files/2019-02/2019_Edelman_Trust_Barometer_Global_Report_2.pdf}
\BIBentrySTDinterwordspacing

\bibitem{tracetogether2020zendesk2}
T.~TraceTogether, ``{How does TraceTogether work?}''
  \url{https://tracetogether.zendesk.com/hc/en-sg/articles/360043543473-How-does-TraceTogether-work-},
  2020, accessed: 2020-03-23.

\bibitem{cdc_spread}
\BIBentryALTinterwordspacing
``How {COVID-19} spreads,'' \emph{Centers for Disease Control and Prevention},
  March 2020. [Online]. Available:
  \url{https://www.cdc.gov/coronavirus/2019-ncov/prepare/transmission.html}
\BIBentrySTDinterwordspacing

\bibitem{van15}
J.~Van Den~Hooff, D.~Lazar, M.~Zaharia, and N.~Zeldovich, ``{Vuvuzela: Scalable
  private messaging resistant to traffic analysis},'' in \emph{Proceedings of
  the 25th Symposium on Operating Systems Principles}, 2015, pp. 137--152.

\bibitem{tyagi2017stadium}
N.~Tyagi, Y.~Gilad, D.~Leung, M.~Zaharia, and N.~Zeldovich, ``Stadium: A
  distributed metadata-private messaging system,'' in \emph{Proceedings of the
  26th Symposium on Operating Systems Principles}, 2017, pp. 423--440.

\bibitem{corrigan15}
H.~Corrigan-Gibbs, D.~Boneh, and D.~Mazi{\`e}res, ``{Riposte: An anonymous
  messaging system handling millions of users},'' in \emph{2015 IEEE Symposium
  on Security and Privacy}.\hskip 1em plus 0.5em minus 0.4em\relax IEEE, 2015,
  pp. 321--338.

\bibitem{freedman2004efficient}
M.~J. Freedman, K.~Nissim, and B.~Pinkas, ``Efficient private matching and set
  intersection,'' in \emph{International conference on the theory and
  applications of cryptographic techniques}.\hskip 1em plus 0.5em minus
  0.4em\relax Springer, 2004, pp. 1--19.

\bibitem{kissner2005privacy}
L.~Kissner and D.~Song, ``Privacy-preserving set operations,'' in \emph{Annual
  International Cryptology Conference}.\hskip 1em plus 0.5em minus 0.4em\relax
  Springer, 2005, pp. 241--257.

\bibitem{de2010practical}
E.~De~Cristofaro and G.~Tsudik, ``Practical private set intersection protocols
  with linear complexity,'' in \emph{International Conference on Financial
  Cryptography and Data Security}.\hskip 1em plus 0.5em minus 0.4em\relax
  Springer, 2010, pp. 143--159.

\bibitem{de2012fast}
E.~De~Cristofaro, P.~Gasti, and G.~Tsudik, ``Fast and private computation of
  cardinality of set intersection and union,'' in \emph{International
  Conference on Cryptology and Network Security}.\hskip 1em plus 0.5em minus
  0.4em\relax Springer, 2012, pp. 218--231.

\bibitem{simmons1979symmetric}
G.~J. Simmons, ``Symmetric and asymmetric encryption,'' \emph{ACM Computing
  Surveys (CSUR)}, vol.~11, no.~4, pp. 305--330, 1979.

\bibitem{greschbach12}
B.~Greschbach, G.~Kreitz, and S.~Buchegger, ``{The devil is in the
  metadata—new privacy challenges in decentralised online social networks},''
  in \emph{2012 IEEE International Conference on Pervasive Computing and
  Communications Workshops}.\hskip 1em plus 0.5em minus 0.4em\relax IEEE, 2012,
  pp. 333--339.

\bibitem{kwon20}
A.~Kwon, D.~Lu, and S.~Devadas, ``{$\{$XRD$\}$: Scalable Messaging System with
  Cryptographic Privacy},'' in \emph{17th $\{$USENIX$\}$ Symposium on Networked
  Systems Design and Implementation ($\{$NSDI$\}$ 20)}, 2020, pp. 759--776.

\bibitem{chaum81}
D.~L. Chaum, ``{Untraceable electronic mail, return addresses, and digital
  pseudonyms},'' \emph{Communications of the ACM}, vol.~24, no.~2, pp. 84--90,
  1981.

\bibitem{reed1998anonymous}
M.~G. Reed, P.~F. Syverson, and D.~M. Goldschlag, ``Anonymous connections and
  onion routing,'' \emph{IEEE Journal on Selected areas in Communications},
  vol.~16, no.~4, pp. 482--494, 1998.

\bibitem{lopez2012fly}
A.~L{\'o}pez-Alt, E.~Tromer, and V.~Vaikuntanathan, ``On-the-fly multiparty
  computation on the cloud via multikey fully homomorphic encryption,'' in
  \emph{Proceedings of the forty-fourth annual ACM symposium on Theory of
  computing}, 2012, pp. 1219--1234.

\bibitem{chen2019efficient}
H.~Chen, W.~Dai, M.~Kim, and Y.~Song, ``Efficient multi-key homomorphic
  encryption with packed ciphertexts with application to oblivious neural
  network inference,'' in \emph{Proceedings of the 2019 ACM SIGSAC Conference
  on Computer and Communications Security}, 2019, pp. 395--412.

\bibitem{news_opensource}
\BIBentryALTinterwordspacing
J.~Zhang, ``{620,000 people installed TraceTogether in 3 days, S’pore’s
  open source contact tracing app},'' \emph{Mothership}, March 2020. [Online].
  Available:
  \url{https://mothership.sg/2020/03/tracetogether-installed-open-source/}
\BIBentrySTDinterwordspacing

\bibitem{wagenaar1995methods}
A.~C. Wagenaar, T.~S. Zobeck, G.~D. Williams, and R.~Hingson, ``{Methods used
  in studies of drink-drive control efforts: a meta-analysis of the literature
  from 1960 to 1991},'' \emph{Accident Analysis \& Prevention}, vol.~27, no.~3,
  pp. 307--316, 1995.

\bibitem{pei2018forecasting}
S.~Pei, S.~Kandula, W.~Yang, and J.~Shaman, ``Forecasting the spatial
  transmission of influenza in the {United States},'' \emph{Proceedings of the
  National Academy of Sciences}, vol. 115, no.~11, pp. 2752--2757, 2018.

\bibitem{sarma2002rfid}
S.~E. Sarma, S.~A. Weis, and D.~W. Engels, ``{RFID} systems and security and
  privacy implications,'' in \emph{International Workshop on Cryptographic
  Hardware and Embedded Systems}.\hskip 1em plus 0.5em minus 0.4em\relax
  Springer, 2002, pp. 454--469.

\bibitem{safekit}
``{Private Kit: Safe Paths- Can we slow the spread without giving up individual
  privacy?}'' \url{http://safepaths.mit.edu/}, 2020, accessed: 2020-03-23.

\bibitem{covidwatch}
``{COVID Watch},'' \url{https://covid-watch.org/}, 2020.

\bibitem{csct}
J.~Petrie, ``{Cryptographically Secure Contact Tracing},'' March 2020.

\bibitem{latimes_limitedtesting}
\BIBentryALTinterwordspacing
J.~Dolan and B.~Mejia, ``{L.A. County gives up on containing coronavirus, tells
  doctors to skip testing of some patients},'' \emph{Los Angeles Times}, March
  2020. [Online]. Available:
  \url{https://www.latimes.com/california/story/2020-03-20/coronavirus-county-doctors-containment-testing}
\BIBentrySTDinterwordspacing

\bibitem{philinq_limitedtesting}
\BIBentryALTinterwordspacing
C.~Y. Johnson and L.~H. Sun, ``{Health officials in New York, California
  restrict coronavirus testing to health care workers and people who are
  hospitalized},'' \emph{The Philadelphia Inquirer}, March 2020. [Online].
  Available:
  \url{https://www.inquirer.com/health/coronavirus/coronavirus-testing-20200321.html}
\BIBentrySTDinterwordspacing

\end{thebibliography}

\end{document}